\documentclass[nofootinbib,preprint,preprintnumbers,a4paper,11pt]{revtex4}
\usepackage{epsfig}
\usepackage{footnote}
\usepackage{ulem}
\usepackage{color}
\usepackage{array}
\usepackage{amssymb}
\usepackage{amsmath}
\usepackage{graphicx}
\usepackage{subfigure}
\usepackage{longtable}
\usepackage{verbatim}
\usepackage{amsfonts}
\usepackage{hyperref}
\usepackage{multirow}
\usepackage{cancel}

\begin{document}

\title{The factorization-assisted topological-amplitude approach and its applications}

\author{Qin Qin$^{1}$\footnote{qqin@hust.edu.cn}, Chao Wang$^{2}$\footnote{chaowang@nankai.edu.cn}, Di Wang$^{3}$\footnote{wangdi@hunnu.edu.cn}, 
Si-Hong Zhou$^{4}$\footnote{shzhou@imu.edu.cn}}

\address{$^1$School of Physics, Huazhong University of Science and Technology, Wuhan 430074, China}
\address{$^2$Faculty of Mathematics and Physics, Huaiin Institute of Technology, Huaian, Jiangsu 223001,
China}
\address{$^3$Department of Physics, Hunan Normal University, Changsha 410081, China}
\address{$^4$School of Physical Science and Technology, Inner Mongolia University, Hohhot 010021, China}

\begin{abstract}
Heavy meson decays provide an important platform for studies of both QCD and electroweak dynamics, which may contain some portals to understanding of nonperturbative QCD and physics beyond the Standard Model.
The factorization-assisted topological-amplitude approach was proposed to study two-body non-leptonic $D$ meson decays, where a promising QCD inspired approach from first principles is still missing.
It was also applied to $B$ meson decays whose subleading power contributions are difficult to calculate.
By factorizing topological amplitudes into short distance Wilson coefficients and long distance hadronic matrix elements either to be calculated or to be parameterized, it provides an effective framework to extract information of nonperturbative dynamics involved.
With important flavor SU(3) breaking effects taken into account, the data of the decay branching ratios (and also CP asymmetries in $B$ decays) can be fitted well.
The extracted amplitudes were further applied to make predictions for other observables, such as CP asymmetries in $D$ decays, mixing parameters in the $D^0-\bar{D}^0$ system, and so on. By this review, we will describe the formulation of the factorization-assisted topological-amplitude approach and summarize its applications in $D$ and $B$ meson decays and highlight some of its achievements.
\end{abstract}

\maketitle

\section{Introduction}

Non-leptonic heavy meson decay provides an ideal platform to test the Standard Model (SM) and search for new physics.
Abundant data of the $B$ and $D$ meson decays have been collected by BaBar, Belle (II), BESIII and LHCb experiments in the
last couple of decades \cite{ParticleDataGroup:2020ssz}.
Just recently, the LHCb Collaboration observed the CP violation in the charm sector with $5.3\sigma$ in 2019 \cite{LHCb:2019hro},
which is another milestone event in the heavy quark physics.
In theory, QCD inspired approaches based on heavy quark expansion have been developed to calculate the non-leptonic $B$ meson decays,
such as QCD factorization (QCDF) \cite{Beneke:1999br,Beneke:2003zv,Beneke:2001ev}, perturbative QCD approach (PQCD) \cite{Keum:2000ph,Keum:2000wi,Lu:2000em,Lu:2000hj}, and soft-collinear effective theory (SCET)~\cite{Bauer:2001cu,Bauer:2001yt,Beneke:2002ph}.
Although significant progresses have been made within these approaches, systematic calculation of
$(\Lambda_{\rm QCD}/m_b)^n$ power corrections is still challenging for all of them.
For $D$ meson decays, the situation is even worse because of the poor convergence with the large expansion parameters
$\alpha_s(m_c)$ and $\Lambda_{\rm QCD}/m_c$. As a result, it is difficult to calculate the $D$ decay amplitudes from first principles of QCD.

One idea is to extract the dynamics of $D$ meson decays from data based on some model assumptions.
One of the popular methods is the topological diagram approach \cite{Cheng:2012xb,Cheng:2012wr,Cheng:2010ry,Chau:1987tk,Chau:1986jb,Cheng:2014rfa,Chau:1990ay,Cheng:2016ejf}.
In this approach, the topological diagrams are classified according to the topologies in the flavor flow of weak decay diagrams and all strong interaction effects induced implicitly.
It provides a framework to analyze nonperturbative effects by data fitting.
However, the usual global fitting is based on the flavor symmetric topological diagrams, with the flavor SU(3) breaking effects not included.
The SU(3) breaking effects should be around 20\% and turn out to be nonnegligible even in $B$ meson decays \cite{Huber:2021cgk}.
It leads to the inability of interpreting the branching fractions, especially in the singly Cabibbo-suppressed $D$ decays.
The topological diagram approach can include the flavor symmetry breaking effects by considering a linear SU(3) breaking \cite{Muller:2015lua},
but it will induce too many parameters to fit limited data which causes very large uncertainties.

The factorization-assisted topological-amplitude approach (FAT) was proposed in
\cite{Li:2012cfa,Qin:2013tje,Zhou:2015jba,Zhou:2016jkv} parameterizing the nonperturbative contributions in the two-body non-leptonic $D/B$ decays,
trying to decode the involved SU(3) breaking effects.
In this framework, the factorization hypothesis is used to calculate topological amplitudes that are factorized into two parts: the short-distance Wilson coefficients and the long-distance hadronic matrix elements. The nonperturbative and nonfactorizable contributions are determined by the experimental data.
It is expected that the dominant SU(3) breaking effects are effectively summarized by several parameters.
It turns out that the FAT can fit the data very well, including those of the $D/B$ decays into two pseudoscalar mesons ($PP$),
one pseudoscalar meson and one vector meson $(PV)$, and two vector mesons ($VV$). Furthermore, the extracted amplitudes information
can be used to calculate other observables. For example, the FAT prediction on the $CP$ asymmetry difference
$\Delta A_{CP}\equiv A_{CP}(D^{0}\to K^{+}K^{-}) -A_{CP}(D^{0}\to \pi^{+}\pi^{-})$ was given in 2012 \cite{Li:2012cfa}, which was recently confirmed by the LHCb
measurement \cite{LHCb:2019hro}.
Besides, the FAT approach can also be applied to study other topics such as neutral $D$ mixing, $K^0_S-K^0_L$ asymmetry, and so on.
In this review, we will briefly introduce the basic ideas of FAT approach and highlight the achievements it has achieved.

The rest of this paper is organized as follows.
In Section~\ref{br}, we review the basic ideas and the results of the FAT approach in the branching fractions of $D$ meson decays.
In Section~\ref{cpv}, we discuss the CP violation and related topics in the charm sector.
In Section~\ref{b}, the application of FAT approach in the $B$ meson decays is reviewed.
Section~\ref{sum} contains a summary and outlook.

\section{Branching fractions of charm decays}\label{br}

\begin{figure}[tph!]
  \centering
  \includegraphics[width=12cm]{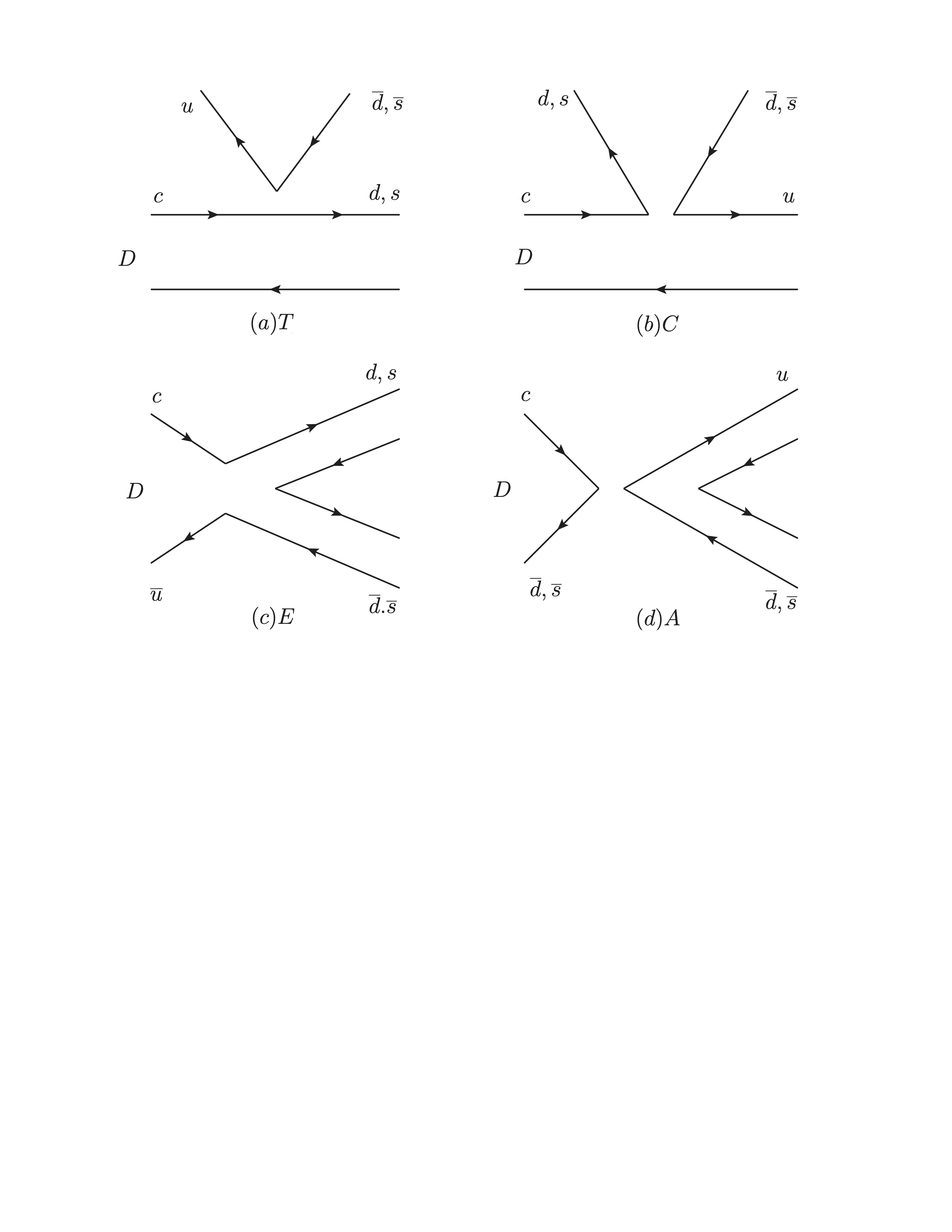}
  \vspace{-8cm}
  \caption{Topological tree diagrams contributing to the two-body non-leptonic $D$ decays with (a) the color-favored penguin amplitude $T$, (b) the color-suppressed penguin amplitude $C$, (c) the gluon-annihilation penguin amplitude $E$, and (d) the gluon-exchange penguin amplitude $A$. The diagrams are taken from \cite{Li:2012cfa}.}\label{tree}
\end{figure}

According to the quark transition of the weak operators, there are four types of amplitudes contributing to the two-body $D$ decays --- the color-favored tree amplitude $T$, the color-suppressed amplitude $C$, the $W$-exchange amplitude $E$ and the $W$-annihilation amplitude $A$, as displayed as the corresponding four topological diagrams in Fig. \ref{tree}.
In the FAT approach, the two-body non-leptonic $D$ meson decays are formulated in a way that the short- and long-distance dynamics are separated with the former calculable and the latter parameterized to be determined by data.
The effective Hamiltonian of charm decay in the SM can be written as
\begin{align}\label{Heff12}
&\mathcal{H}_{eff}=\frac{G_F}{\sqrt{2}}
 V_{CKM}\left[C_1(\mu)Q_1(\mu)+C_2(\mu)Q_2(\mu)\right]+H.c.,\nonumber\\
& Q_1=\bar{u}_{\alpha}\gamma_\mu(1-\gamma_5)q_{2\beta}
\bar{q}_{1\beta}\gamma^\mu(1-\gamma_5)c_{\alpha},~~~~~
Q_2=\bar{u}_{\alpha}\gamma_\mu(1-\gamma_5)q_{2\alpha}
\bar{q}_{1\beta}\gamma^\mu(1-\gamma_5)c_{\beta},
\end{align}
where $G_F$ denotes the Fermi coupling constant, $V_{\rm CKM}$ is the products of the Cabibbo-Kobayashi-Maskawa (CKM) matrix elements, $C_{1,2}$ are the Wilson coefficients, $\alpha,\beta$ are the color indices.
The physics above the scale $\mu\sim m_c$ is handled by the Wilson coefficients $C_{1,2}(\mu)$, and the physics below $\mu$
is handled by the hadronic matrix elements of the current-current operators $\langle M_1M_2 | O_{1,2} | D\rangle$.
In the FAT approach, the factorizable parts of the four amplitudes are either calculated by the naive factorization as the
products of the corresponding decay constants and form factors, or neglected
owing to the color suppression or the helicity suppression. The non-factorizable parts are parameterized with non-perturbative parameters and extracted by fitting.

Taking the $D\to PP$ decay as an example, the topological amplitudes in the FAT are \cite{Li:2012cfa}
\begin{align}
T[C] &=  \frac{G_f}{\sqrt{2}}V_{CKM}a_{1}(\mu)[a_{2}(\mu)]f_{P_2}(m^2_D - m^2_{P_1})F_0^{D\rightarrow P_1}(m^2_{P_2}),\label{eq:TPP}
\\
E &=  \frac{G_f}{\sqrt{2}}V_{CKM}C_2(\mu)\chi^E_{q,s}e^{i\phi^E_{q,s}}f_D m^2_D \Big(\frac{f_{P_1}f_{P_2}}{f_\pi^2}\Big),\label{eq:EPP}
\\
A &=  \frac{G_f}{\sqrt{2}}V_{CKM}C_1(\mu)\chi^A_{q,s}e^{i\phi^A_{q,s}}f_D m^2_D \Big(\frac{f_{P_1}f_{P_2}}{f_\pi^2}\Big),\label{eq:APP}
\end{align}
with
\begin{align}
a_1(\mu) &= C_2(\mu) + \frac{C_1(\mu)}{N_c}, \quad
a_2(\mu) = C_1(\mu) + C_2(\mu)[\frac{1}{N_c} + \chi^{C}e^{i\phi^{C}}].
\end{align}
In $T$ and $C$ diagrams, $P_1$ is the pseudoscalar meson transited from the $D$ decays and $P_2$ the emitted meson. $T$ diagram is calculated in the factorization hypothesis, $f_{i}$ and $F_0$ are the decay constants and transition form factors, respectively.
The nonfactorizable contributions in the $C$ diagram, resulting from the final-state interactions, are parametrized as magnitude and strong phase as $\chi^{C} e^{i\phi^{C}}$.
$E$ and $A$ diagrams are dominated by the nonfactorizable contributions, parametrized as $\chi_{q,s}^{E,A}e^{i\phi_{q,s}^{E,A}}$.
The subscripts $q$ and $s$ stand for the quark pairs produced from the vacuum as the $u$, $d$ quarks or the $s$ quark.
Because of the fact that the pion boson is a Nambu-Goldstone boson and quark-antiquark bound state simultaneously, a Glauber phase factor \cite{Li:2014haa,Li:2021req}, $e^{iS_\pi}$, is introduced for each pion involved in the non-factorizable contributions of $E$ and $A$ amplitudes.
The FAT formulation covers the flavor SU(3) breaking effects in several aspects, by different meson decay constants and transition form factors, by different phase spaces, by different scale choices, and by the Glauber gluon effects only significant
in the pion involved channels. One can refer to \cite{Li:2012cfa,Qin:2013tje} for more details.

\begin{table}[tph!]
\caption{Comparison for FAT predictions and experimental data of branching ratios of the $D\to PP$ decays in units of permill. Only
the experimental results published after the theoretical predictions are listed.}\label{tab:charmpp}
\begin{tabular}{|c|c|c|}
\hline\hline
{\qquad Channel \qquad } & {\qquad FAT \qquad~ }& {\qquad data \qquad}   \\\hline
$D^0 \to K_S^0 K_S^0$ & $0.073\pm 0.002$~\cite{Jiang:2017zwr} & $0.167\pm 0.011\pm 0.011$~\cite{BESIII:2016nrs} \\ \hline
$D^0 \to \pi^0 \eta$ & $0.74\pm 0.03$~\cite{Jiang:2017zwr} & $0.58\pm 0.05\pm 0.05$~\cite{BESIII:2018oqs} \\ \hline
$D^0 \to \pi^0 \eta^\prime$ & $1.08\pm 0.05$~\cite{Jiang:2017zwr} & $0.93\pm 0.11\pm 0.09$~\cite{BESIII:2018oqs} \\ \hline
$D^0 \to \eta \eta$ & $1.86\pm 0.06$~\cite{Jiang:2017zwr} & $2.20\pm 0.07\pm 0.06$~\cite{BESIII:2018oqs} \\ \hline
$D^0 \to \eta \eta^\prime$ & $1.05\pm 0.08$~\cite{Jiang:2017zwr} & $0.94\pm 0.25\pm 0.11$~\cite{BESIII:2018oqs} \\ \hline
$D^0 \to \pi^+ \pi^-$ & $1.44\pm 0.02$~\cite{Jiang:2017zwr} & $1.508\pm 0.018\pm 0.022$~\cite{BESIII:2018apz} \\ \hline
$D^0 \to K^+ K^-$ & $4.05 \pm 0.07 $~\cite{Jiang:2017zwr} & $4.233 \pm 0.021 \pm 0.064 $~\cite{BESIII:2018apz} \\ \hline
$D^0 \to K^\mp \pi^\pm$ & $39.3 \pm 0.4$~\cite{Jiang:2017zwr} & $38.98 \pm 0.06 \pm 0.51$~\cite{BESIII:2018apz} \\ \hline
$D^0 \to K_S^0 \pi^0$ & $12.1 \pm 0.4$~\cite{Jiang:2017zwr} & $12.39 \pm 0.06 \pm 0.27$~\cite{BESIII:2018apz} \\ \hline
$D^0 \to K_S^0 \eta$ & $4.8 \pm 0.3$~\cite{Jiang:2017zwr} & $5.13 \pm 0.07 \pm 0.12$~\cite{BESIII:2018apz} \\ \hline
$D^0 \to K_S^0 \eta^\prime$ & $9.8 \pm 0.5$~\cite{Jiang:2017zwr} & $9.49 \pm 0.20 \pm 0.36$~\cite{BESIII:2018apz} \\ \hline
$D^+ \to \pi^+\pi^0 $ & 0.89~\cite{Li:2012cfa} & $ 1.259 \pm 0.033 \pm 0.023$~\cite{BESIII:2018apz} \\ \hline
$D^+ \to K^+\pi^0 $ & 0.197~\cite{Li:2012cfa} & $  0.231 \pm 0.021 \pm 0.006$~\cite{BESIII:2018apz} \\ \hline
$D^+ \to \pi^+\eta $ & 3.39~\cite{Li:2012cfa} & $  3.790 \pm 0.070 \pm0.068 $~\cite{BESIII:2018apz} \\ \hline
$D^+ \to K^+\eta $ & 0.066~\cite{Li:2012cfa} & $   0.151 \pm 0.025 \pm 0.0148 $~\cite{BESIII:2018apz} \\ \hline
$D^+ \to \pi^+\eta^\prime $ & 4.58~\cite{Li:2012cfa} & $  5.12 \pm 0.14 \pm 0.21 $~\cite{BESIII:2018apz} \\ \hline
$D^+ \to K^+\eta^\prime $ & 0.114~\cite{Li:2012cfa} & $  0.164 \pm 0.051 \pm 0.024 $~\cite{BESIII:2018apz} \\ \hline
$D^+ \to \pi^+K_S^0 $ & 16.2~\cite{Li:2012cfa} & $ 15.91 \pm 0.06 \pm 0.30 $~\cite{BESIII:2018apz} \\ \hline
$D^+ \to K^+ K_S^0 $ & 2.98~\cite{Li:2012cfa} & $  3.183 \pm 0.029 \pm 0.060 $~\cite{BESIII:2018apz} \\ \hline
$D_s^+ \to \pi^+ \pi^0$ & 0~\cite{Li:2012cfa} & $0.037 \pm 0.055 \pm 0.021 \pm 0.001 $~\cite{Belle:2021ygw} \\ \hline
$D_s^+ \to \pi^0 K^+$ & 0.67~\cite{Li:2012cfa} & $0.748 \pm 0.049 \pm 0.018 \pm 0.023 $~\cite{BESIII:2020kim}; $0.735 \pm 0.052 \pm 0.030 \pm 0.026 $~\cite{Belle:2021ygw} \\ \hline
$D_s^+ \to \pi^+K_S^0$ & 1.105~\cite{Li:2012cfa} & $1.109 \pm 0.034 \pm 0.023 \pm 0.035 $~\cite{BESIII:2020kim} \\ \hline
$D_s^+ \to K^+K_S^0$ & 15.03~\cite{Li:2012cfa} & $ 15.02 \pm 0.10 \pm 0.27 \pm 0.47 $~\cite{BESIII:2020kim} \\ \hline
$D_s^+ \to \pi^+\eta $ & 16.5~\cite{Li:2012cfa} & $ 17.41 \pm 0.18 \pm 0.27 \pm 0.54 $~\cite{BESIII:2020kim}; $19.00 \pm 0.10 \pm 0.59 \pm 0.68 $~\cite{Belle:2021ygw} \\ \hline
$D_s^+ \to K^+\eta $ & 1.0~\cite{Li:2012cfa} & $ 1.62 \pm 0.10 \pm 0.03 \pm 0.05 $~\cite{BESIII:2020kim};  $1.75\pm 0.05 \pm 0.05 \pm 0.06$~\cite{Belle:2021ygw}  \\ \hline
$D_s^+ \to \pi^+\eta^\prime $ & 34.4~\cite{Li:2012cfa} & $ 37.8 \pm 0.4 \pm 2.1 \pm 1.2 $~\cite{BESIII:2020kim} \\ \hline
$D_s^+ \to K^+\eta^\prime $ & 1.92~\cite{Li:2012cfa} & $ 2.68 \pm 0.17 \pm 0.17 \pm 0.08 $~\cite{BESIII:2020kim} \\ \hline
\hline
\end{tabular}
\end{table}

\begin{table}[tph!]
\caption{Comparison for FAT predictions and experimental data of branching ratios of the $D\to PV$ decays in units of permill. Only
the experimental results published after the theoretical predictions are listed.}\label{tab:charmpv}
\begin{tabular}{|c|c|c|}
\hline\hline
{\qquad Channel \qquad } & {\qquad FAT \qquad~ }& {\qquad data \qquad}  \\\hline
$D_s^+\to \eta^\prime \rho^+$ & 1.7(1.6)~\cite{Qin:2013tje} & $58\pm 14\pm 4$~\cite{BESIII:2015rrp} \\ \hline
$D^0 \to \eta \omega$ & $2.1\pm 0.1$~\cite{Jiang:2017zwr} & $2.15\pm 0.17\pm 0.15$~\cite{BESIII:2018oqs};  $1.78 \pm 0.19 \pm0.14$~\cite{Smith:2018ryf} \\ \hline
$D_s^+\to \pi^+ \omega$ & 3.0(2.6)~\cite{Qin:2013tje} & $1.77\pm 0.32\pm 0.11$ ~\cite{BESIII:2018mwk} \\ \hline
$D_s^+\to K^+ \omega$ & 0.6(0.07)~\cite{Qin:2013tje} & $0.87\pm 0.24\pm 0.07$ ~\cite{BESIII:2018mwk} \\ \hline
$D^0 \to \pi^0 \phi$ & $1.4\pm 0.1$~\cite{Jiang:2017zwr} & $1.168\pm 0.028\pm 0.028$~\cite{1907.11258} \\ \hline
$D^0 \to \eta \phi$ & $0.18\pm 0.04$~\cite{Jiang:2017zwr} & $0.181\pm 0.046\pm 0.006$~\cite{1907.11258} \\ \hline
$D^+ \to \pi^+ \phi$ & 5.65(5.65)~\cite{Jiang:2017zwr} & $5.70\pm 0.05\pm 0.13$~\cite{1907.11258} \\ \hline
$D^+ \to K^+ \phi$ & 0.001(0.002)~\cite{Jiang:2017zwr} & $< 0.021$~\cite{1907.11258} \\ \hline
$D^0 \to \eta \bar{K}^{*0}$ & $6.1\pm 1.0$~\cite{Jiang:2017zwr} & $14.1^{+1.3}_{-1.2}$~\cite{Belle:2020fbd}; $17.7\pm4.4$ ~\cite{BESIII:2020pxp} \\ \hline
$D^+ \to K^+ \omega$ & 0.09(0.07)~\cite{Jiang:2017zwr} & $0.057^{+0.025}_{-0.021}\pm0.002$~\cite{BESIII:2020wnc} \\ \hline
$D^+ \to K^0_S K^{*+}$ & 12.56(15.56)~\cite{Jiang:2017zwr} & $8.69 \pm 0.40 \pm 0.64 \pm 0.51$~\cite{BESIII:2021dmo} \\ \hline
$D^0 \to \eta\phi$ & $0.18\pm0.04$~\cite{Jiang:2017zwr} & $0.184 \pm 0.009 \pm 0.006 \pm 0.005$~\cite{Belle:2021dfa} \\ \hline\hline
$D^0 \to \omega\phi$ & & $6.48 \pm 0.96 \pm 0.38)\times 10^{-4}$ (Transverse)~\cite{BESIII:2021raf} \\ \hline
\end{tabular}
\end{table}

The fits to the data of the branching ratios for the $PP$ and $PV$ channels are performed in \cite{Li:2012cfa} and \cite{Qin:2013tje}, respectively.
An update was made by \cite{Jiang:2017zwr} for the neutral $D^0$ meson decays. The consistency between the fitted results and
the experimental data indicates that both the perturbative and non-perturbative dynamics of the two-body non-leptonic $D$ meson
decays, especially the flavor SU(3) breaking effects, are well handled by the FAT approach. The non-factorizable magnitude parameters,
which should be power suppressed in the heavy
quark expansion, turn out to be not so small compared to the factorizable ones. This confirms our expectation that the heavy quark expansion
does not work well for $D$ meson decays, which is one of the main motivations of the FAT approach.

One important achievement of the FAT approach is that it provided a solution to the long-standing $\pi\pi$-$KK$ puzzle ---
the measured $D^0\to K^+K^-$ branching ratio is about three times the $D^0\to \pi^+\pi^-$ branching ratio, while the previous
theoretical calculations predicted a smaller branching ratio for $D^0\to K^+K^-$. As pointed out in \cite{Cheng:2012wr},
the $W$-exchange amplitudes in the two channels must have not only different magnitudes but also different strong phases
to give the correct branching ratios. In the FAT approach, different decay constants and form factors can account for different
magnitudes. More importantly, the Glauber gluon effects involved in the $\pi^+\pi^-$ channel appropriately rotate the phase angle in the
$E$ amplitude to give a correct prediction for the $\pi^+\pi^-$ branching ratio and the correct relation between the $\pi\pi$-$KK$ channels.
This is another evidence that the SU(3) breaking effects are well under control in the FAT approach, which is crucial to the study of the
$D^0-\bar{D}^0$ mixing, which will be discussed in the next section.

In the Global fit of the $PV$ channels \cite{Qin:2013tje}, there was one channel $D_s^+\to\rho^+\omega$ whose FAT fit result was never
consistent with the PDG value $\mathcal{B}(D_s^+\to\rho^+\omega)=(11.7\pm1.8)\%$ at that time \cite{ParticleDataGroup:2012pjm}.
The FAT prediction $1.6\%-1.7\%$ is more consistent with a new measurement by the CLEO-c, $(5.6\pm1.1)\%$ \cite{CLEO:2013bae},
which was later confirmed by a BESIII measurement $(5.8\pm 1.4\pm 0.4)$\% \cite{BESIII:2015rrp}.

After the publications of \cite{Li:2012cfa,Qin:2013tje,Jiang:2017zwr}, the branching ratios of many observed $D$ meson decay channels
have been remeasured with higher precisions and also many new $D$ meson decay channels have been discovered by
experiments. We list their results in TABLE \ref{tab:charmpp} and \ref{tab:charmpv} for the $PP$ and $PV$ channels, respectively,
compared to the FAT predictions. They are basically in consistence with each other.

\section{CP violation and related topics in charm}\label{cpv}

\begin{figure}[tph!]
  \centering
  \includegraphics[width=10cm]{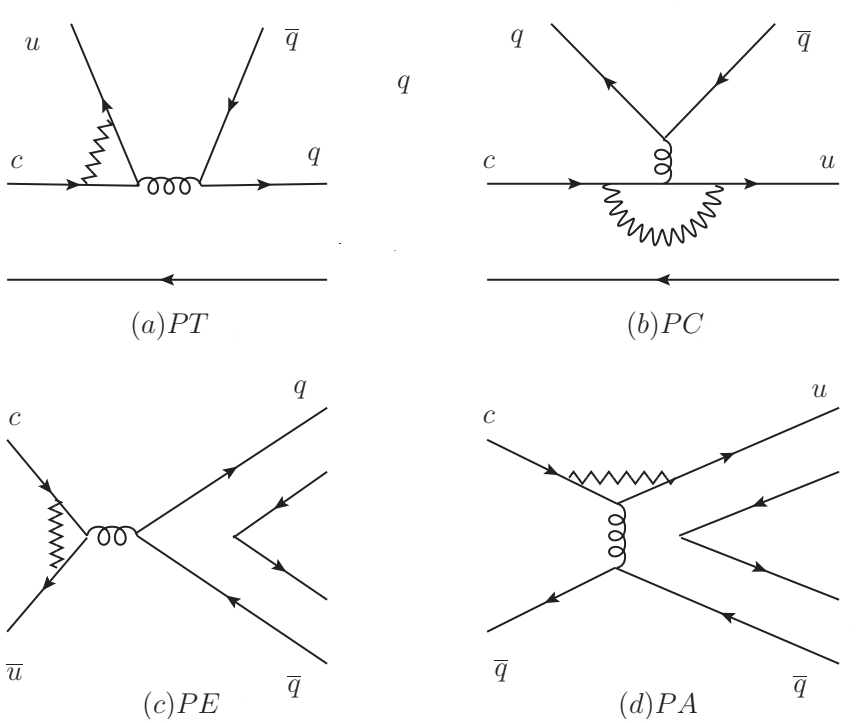}
  \caption{Topological penguin diagrams contributing to the two-body non-leptonic $D$ decays with (a) the color-favored penguin amplitude $PT$, (b) the color-suppressed penguin amplitude $PC$, (c) the gluon-annihilation penguin amplitude $PE$, and (d) the gluon-exchange penguin amplitude $PA$. The diagrams are taken from~\cite{Qin:2013tje}.}\label{penguin}
\end{figure}

Except for predictions of the branching fractions, the FAT approach also provides a prescription that the penguin amplitudes can be related to the tree amplitudes and then the CP asymmetries in the $D$ meson decays could be predicted. Four topological penguin diagrams contributing to the two-body non-leptonic $D$ decays are shown in Fig.~\ref{penguin}.
In the FAT approach, penguin amplitudes $P_T$ and $P_C$ as well as the nonfactorizable contributions to $P_E$ and $P_A$ are estimated by replacement of the Wilson coefficients of the corresponding tree amplitudes in which the nonperturbative parameters have been determined by fitting branching ratios.
The penguin amplitudes that do not related to tree amplitudes are either factorizable and estimated in the pole model, or suppressed by the helicity conservation and negligible. More details can be found in \cite{Li:2012cfa,Qin:2013tje}.
The penguin amplitudes are enhanced by the nonperturbative-QCD effects in the FAT approach, $\mathcal{P}/\mathcal{T}=\mathcal{O}(1)$, which results in a $10^{-3}$ order CPV in the $D$ meson decays compared to the naive expectation that
  $A_{CP}^{\rm charm}\sim\frac{\alpha_s(\mu_c)}{\pi}\frac{|V_{ub}V_{cb}^*|}{|V_{us}V_{cs}^*|} = \mathcal{O}(10^{-4})$.

\begin{figure}[tph!]
  \centering
  \includegraphics[width=6cm]{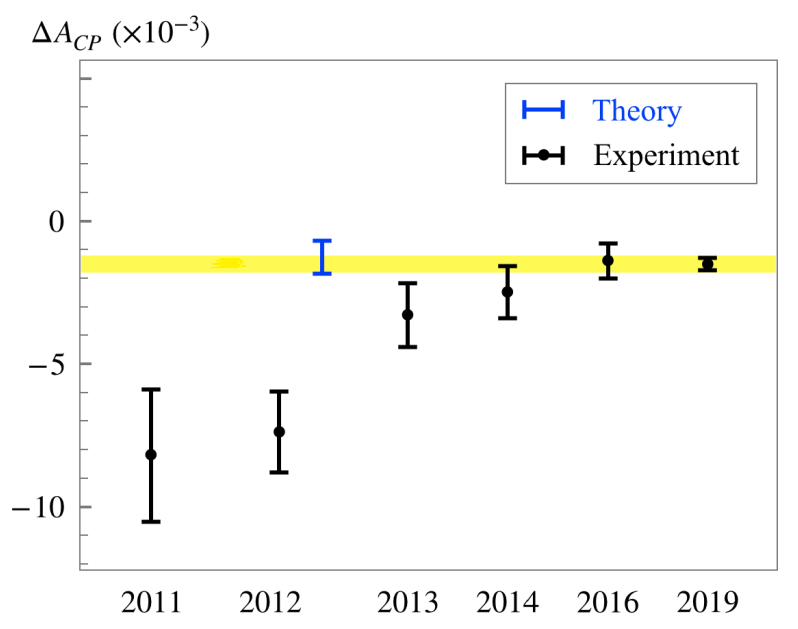}
  \caption{Comparison of $\Delta A_{CP}$ between experimental measurements (in black) and theoretical prediction of the FAT approach (in blue). Experimental results are corresponding to the world-average values for specific year as extracted by the HFLAV~\cite{HFLAV:2019otj}. The yellow band is the most recent experimental result for comparison. The picture is taken from~\cite{Saur:2020rgd}.}\label{par}
\end{figure}
Precise prediction of the difference of CP asymmetries in the $D^0\to K^+K^-$ and $D^0\to \pi^+\pi^-$ modes (known as $\Delta A_{CP}$) is a remarkable achievement of the FAT approach.
As early as 2011, the LHCb Collaboration reported an evidence of charm CPV with $\Delta A_{CP} = (-8.2\pm 2.4)\times 10^{-3}$ \cite{LHCb:2011osy}.
Because of the different understanding of the nonperturbative dynamics in penguin, the theoretical estimation for $\Delta A_{CP}$ ranging from $10^{-4}$ to $10^{-2}$. In 2012, the FAT approach predicted that
\begin{align}
\Delta A_{CP}^{\rm FAT}=(-1.87\sim -0.57)\times 10^{-3}.
\end{align}
It is much smaller than the data in 2011 but consistent with the newest result by the LHCb measurement~\cite{LHCb:2019hro}:
\begin{align}
 \Delta A_{CP}^{\rm EXP} = (-1.54\pm 0.29)\times 10^{-3}.
\end{align}
A comparison between the experimental measurements and the FAT prediction of $\Delta A_{CP}$ is shown in Fig.~\ref{par}.
The consistency of the experimental result with the theoretical prediction indicates reliability of the FAT approach for estimating the penguin amplitudes.
Although some theorists proposed the beyond SM explanations for the large observed CPV \cite{Chala:2019fdb,Dery:2019ysp,Calibbi:2019bay},
we prefer to believe it comes from the non-perturbative QCD enhancement.
Another work that gave a reasonable prediction of $\Delta A_{CP}$ in 2012 is Ref.~\cite{Cheng:2012xb}, in which the penguin-exchange diagram
is assumed to be identical to the W-exchange diagram, $PE=E$. $\Delta A_{CP}$ is predicted to be $(-1.39\pm 0.04)\times 10^{-3}$ or $(-1.51\pm 0.04)\times 10^{-3}$, and updated with similar results in \cite{Cheng:2019ggx}. However, the theoretical explanation of $PE=E$ was not given in \cite{Cheng:2012xb,Cheng:2019ggx}. The reliability and potential dynamics of this hypothesis should be further studied.

\begin{table}[htp!]
\caption{Comparison for FAT predictions and experimental data of CP asymmetries in the singly Cabibbo-suppressed charm decays in units of $10^{-3}$.
}\label{tab:table2}
\begin{tabular}{|c|c|c|}
\hline\hline
{\qquad Channel \qquad } & {\qquad Theoretical prediction \qquad~ }& {\qquad Experimental result \qquad}  \\\hline
$D^0\to \pi^+  \pi^-  $  & $0.58$ & $1.2\pm 1.4$~\cite{HFLAV:2019otj}\\\hline
$D^0\to \pi^0  \pi^0  $  & $0.05$ & $-0.3\pm 6.4$~\cite{HFLAV:2019otj}\\\hline
$D^0\to K^0  \overline K^0  $  & $1.38$ & $-19\pm 10$~\cite{HFLAV:2019otj}\\\hline
$D^0\to K^+  K^-  $  & $-0.42$ & $-0.9\pm 1.1$~\cite{HFLAV:2019otj}\\\hline
$D^0\to \eta  \phi  $  & $0.003$ & $-19\pm 44\pm6$~\cite{Belle:2021dfa}\\\hline
$D^+\to \pi^+  \pi^0  $  & $0$ & $4\pm 8$~\cite{HFLAV:2019otj};\, $-13\pm 9\pm6$~\cite{LHCb:2021rou}\\\hline
$D^+\to \pi^+  \eta  $  & $-0.26$ & $3\pm 7$~\cite{HFLAV:2019otj};\, $-2\pm 8\pm4$~\cite{LHCb:2021rou}\\\hline
$D^+\to \pi^+  \eta^\prime  $  & $1.18$ & $-6\pm 7$~\cite{HFLAV:2019otj}\\\hline
$D^+\to \pi^+  \phi  $  & $-0.0001$ & $0.03\pm 0.40\pm0.29$~\cite{HFLAV:2019otj}\\\hline
$D^+_s\to K^+  \pi^0  $  & $0.39$ & $20\pm 30$~\cite{HFLAV:2019otj};\, $64\pm 44\pm11$~\cite{Belle:2021ygw};\, $-8\pm 39\pm12$~\cite{LHCb:2021rou}\\\hline
$D^+_s\to K^+  \eta  $  & $0.70$ & $19\pm 19$~\cite{HFLAV:2019otj};\, $21\pm 21\pm4$~\cite{Belle:2021ygw};\, $9\pm 37\pm11$~\cite{LHCb:2021rou}\\\hline
$D^+_s\to K^+  \eta^\prime  $  & $-1.60$ & $60\pm 189\pm 9$~\cite{HFLAV:2019otj}\\\hline\hline
\end{tabular}
\end{table}
In Refs.~\cite{Li:2012cfa,Qin:2013tje}, the CP asymmetries in other $D$ decay channels are calculated in the FAT approach.
To compare the FAT predictions with experimental data, we list them in Table.~\ref{tab:table2}.
Because of the large uncertainties, it is difficult to test the FAT approach with experimental data of individual decay channels at the present stage.
More precise experimental observations are looked forward.
Besides, the three-body decay $D^+\to K^+K^-\pi^+$ might be the next potential mode to reveal the CPV in charm due to the large branching fraction $\mathcal{B}r(D^+\to K^+K^-\pi^-)=(9.51\pm 0.34)\times 10^{-3}$ and the large CPV in the two-body decay $D^+\to K^+\overline K^{*}_0(1430)^0$ according to the FAT approach~\cite{Li:2019hho}.

\begin{figure}[tph!]
  \centering
  \includegraphics[width=4.5cm]{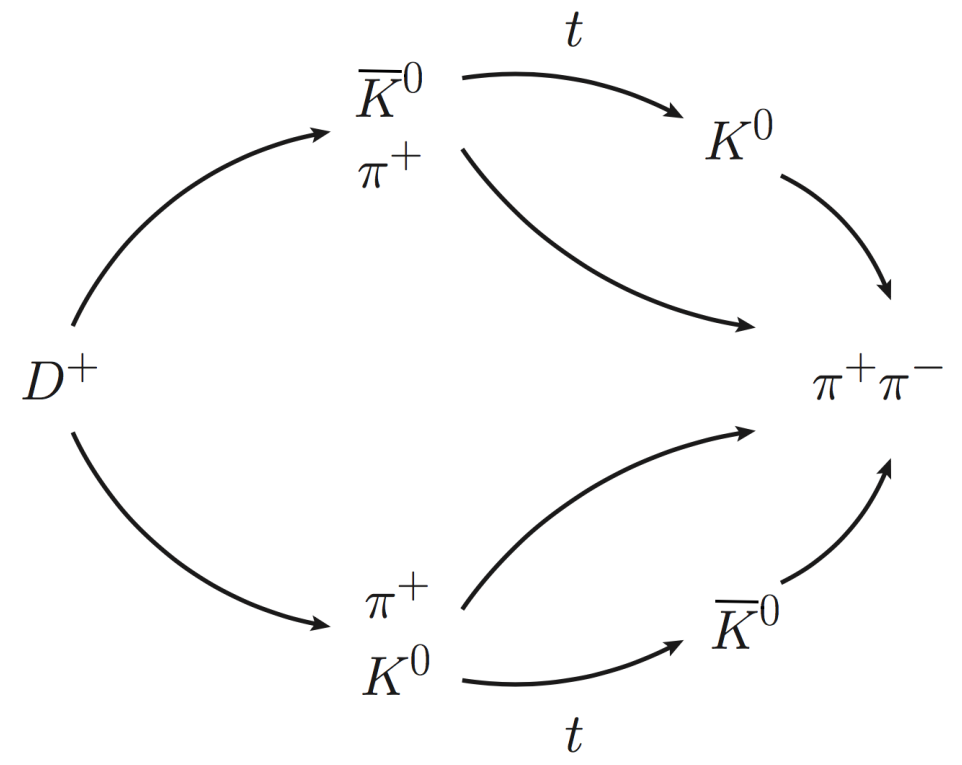}
  \qquad\qquad
    \includegraphics[width=7cm]{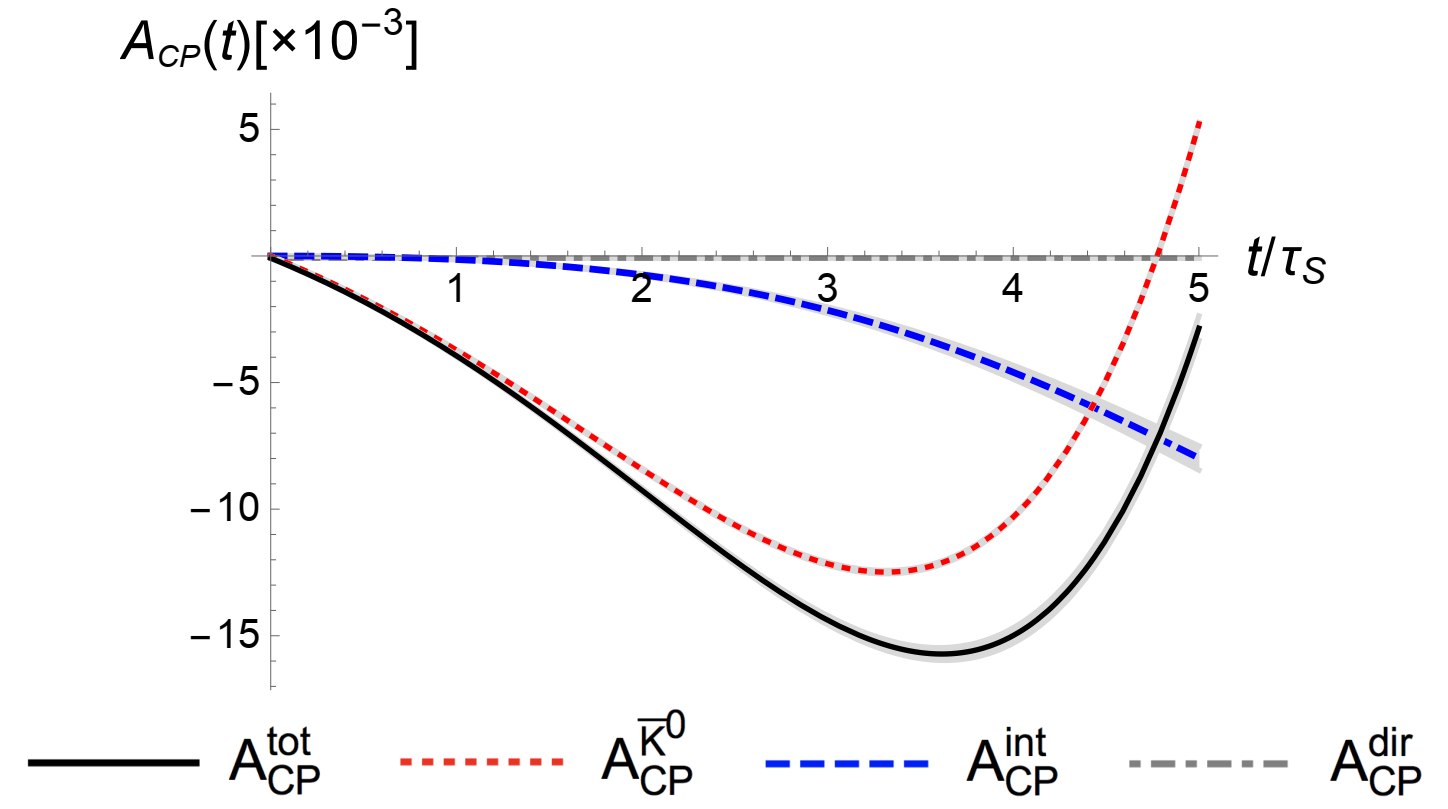}\\
    (a)\qquad\qquad\qquad\qquad\qquad\qquad\qquad\qquad\qquad\qquad (b)~~~~~~~~~~~~
  \caption{Schematic description of the chain decay $D^+\to \pi^+K(t)(\to \pi^+\pi^-)$ (a) and time-dependent CP asymmetries as functions of $t/\tau_S$ (b) taken from~\cite{Yu:2017oky}.}\label{dks}
\end{figure}
Due to the large non-perturbative effects at the charm scale, it is difficult to use the observed CPV in the singly Cabibbo-suppressed $D$ decays to search for new physics.
Compared to the SCS case, both the CF and DCS amplitudes are at the tree level. The amplitudes extracted from branching fractions in the FAT approach are more reliable in the CF and DCS modes and hence could provide a promising signal of new physics. In Ref.~\cite{Yu:2017oky}, we pointed out a new CP-violation effect in charm decays into neutral kaons resulting from the interference between two tree (Cabibbo-favored and doubly Cabibbo-suppressed) amplitudes with the mixing of final-state mesons and predicted it in the FAT approach. A schematic description of the chain decay $D^+\to \pi^+K(t)(\to \pi^+\pi^-)$ and the time-dependent CP asymmetries in this mode as functions of $t/\tau_S$ are shown in Fig.~\ref{dks}.
It is found the new effect $A^{\rm int}_{CP}$ could reach an order of $10^{-3}$ or even $10^{-2}$ in the range of $2\tau_S<t<5\tau_S$, much larger
than the direct CP asymmetry $A^{\rm dir}_{CP}$ induced by interference between the Cabibbo-favored and doubly Cabibbo-suppressed amplitudes.
This discovery corrected the long-standing misunderstanding that only the CPV in $K^0-\overline K^0$ mixing and the direct CPV in charm decay exist in the charm decays into neutral kaons. It is instructive to search for new physics in charm sector.
The new effect is accessible in experiments and could be revealed by measuring the difference of the time-dependent CP asymmetries in the $D^+\to \pi^+K^0_S$ and $D^+\to \pi^+K^0_S$ modes on the LHCb  and Belle II.

\begin{table}[htp!]
\caption{\label{tab:table1}Comparison of $K_S^0-K_L^0$ asymmetries in the $D^0\to K_{S,L}^0\pi^0$, $D^+\to K_{S,L}^0\pi^+$ and $D_s^+\to K_{S,L}^0K^+$ modes estimated in the FAT approach with other methods \cite{Bhattacharya:2009ps,Cheng:2010ry,Gao:2014ena, Muller:2015lua} and the experimental data \cite{CLEO:2007rhw}. The table is taken from~\cite{Wang:2017ksn}. }
\begin{ruledtabular}
\footnotesize\begin{tabular}{ccccccc}
  &$R$\cite{ Bhattacharya:2009ps}&$R$\cite{Cheng:2010ry} & $R$\cite{Gao:2014ena} & $R$\cite{Muller:2015lua} & $R_{\text{exp}}$\cite{CLEO:2007rhw} & $R(\text{FAT})$ \\
  \hline
   $D^0\to K_{S,L}^0\pi^0$ & $0.107$& $0.107$& $0.106$ & $0.09^{+0.04}_{-0.02}$ & $0.108\pm0.035$ & $0.113\pm0.001$ \\
  $D^+\to K_{S,L}^0\pi^+$ &$-0.005\pm0.013$& $-0.019\pm0.016$&$-0.010\pm0.026$ &  & $0.022\pm0.024$ & $0.025\pm0.008$ \\
  $D_s^+\to K_{S,L}^0K^+$ & $-0.002\pm0.009$ &$-0.008\pm0.007$& $-0.008\pm0.007$ & $0.11^{+0.04}_{-0.14}$ &  & $0.012\pm0.006$ \\
\end{tabular}
\end{ruledtabular}
\end{table}
The FAT approach has also been applied to the studies of $K_{S}^{0}-K_{L}^{0}$ asymmetries which induced by interference between the Cabibbo-favored and the doubly Cabibbo-suppressed amplitudes in the $D$ meson decays \cite{Wang:2017ksn}.
The comparison of $K_S^0-K_L^0$ asymmetries of the $D^0\to K_{S,L}^0\pi^0$, $D^+\to K_{S,L}^0\pi^+$ and $D_s^+\to K_{S,L}^0K^+$ modes estimated in the FAT approach with the diagrammatic approach \cite{ Bhattacharya:2009ps,Cheng:2010ry}, the QCD factorization approach \cite{Gao:2014ena}, the diagrammatic approach with global linear SU(3) breaking analysis \cite{Muller:2015lua}, and the experimental data \cite{CLEO:2007rhw} is found in Table~\ref{tab:table1}.
It is found the $K_{S}^{0}-K_{L}^{0}$ asymmetries in the $D^0$-meson decays are shifted by the $D^0-\overline D^0$ mixing parameter $y_{D}\simeq0.006$ in our work.
The result of $R(D^+\to K_{S,L}^{0}\pi^+)$ and $R(D_s^+\to K_{S,L}^0K^+)$ in the FAT approach has the same sign with the experimental data, but opposite to the other theoretical predictions because of the the significant flavor $SU(3)_F$ breaking effects compared to Refs.~\cite{Bhattacharya:2009ps,Cheng:2010ry,Gao:2014ena}.

The FAT approach could also provide us with a glimpse to the charm mixing dynamics since the contributions to $D^0-\overline D^0$ mixing from individual intermediate channels can be summed up in an exclusive approach.
In Ref.~\cite{Jiang:2017zwr}, we estimated the $D^0-\overline D^0$ mixing parameter $y$ in the FAT approach.
Compared to the diagrammatic approach based on the $SU(3)_F$ symmetry \cite{Cheng:2010rv}, the FAT approach provides a more precise treatment of the SU(3) breaking of strong phases.
It is found the contribution from the $PP$ and $PV$ modes is $y_{PP+PV}=(0.21\pm 0.07)\%$, lower than the experimental value $y_{\rm exp}=(0.61\pm 0.08)\%$ \cite{HFLAV:2019otj} but consistent with the fact that $y$ is generated at second order $U$-spin breaking \cite{Gronau:2012kq}.
Since the $U$-spin breaking effects are expected to be more significant in the multi-body $D$ meson decays, they should be included in the evaluation of $y$.
However, it is very difficult to control the $SU(3)_F$ breaking effects in all these modes in an exclusive approach.
A new strategy is necessary to understand the charm mixing dynamics in the Standard Model. And treating it as an inverse problem is an attractive attempt~\cite{Li:2020xrz}.

\section{Bottom decays}\label{b}
In this section, we shall review the application of the FAT approach in the two-body nonleptonic $B$ mesons decays,
including charmed $B$ decays $B\to D^{(*)} P(V)$ \cite{Zhou:2015jba} and charmless $B$ decays $B \to PP,PV$~\cite{Zhou:2016jkv} and $VV$ \cite{Wang:2017rmh}.
The basis idea of the parametrization for both the tree and penguin amplitudes will be briefly discussed, and some selected results for the branching ratios and CP asymmetries
will be presented.

\subsection{Two-body charmed $B$ meson decays}
The charmed hadronic $B$ decay processes have no contribution from penguin operators.
Similar to the $D$ meson decays, four kinds of relevant topologies, $T$, $C$,
$E$ and $A$, are involved in the charmed hadronic decays of $B$ mesons.
In order to keep the SU(3) breaking effects in the decay amplitudes, we factorize the decay constants and form factors formally from each topological amplitude assisted by factorization hypothesis.
And those universal contributions in topological amplitudes are parameterized as some free parameters, $\chi^C, \phi^C, \chi^E\, \text{and}\, \phi^E$.
With the four fitted parameters from 31 decay modes induced by $b\to c$ transition, we then predicted the branching fractions of 120 decay modes induced though both $b\to c$ and $b\to u$ transitions which can be found in the tables of branching ratios in \cite{Zhou:2015jba}.
Our results are well consistent with the measured data
or to be tested in the LHCb and Belle-II experiments in the future.
We found the SU(3) symmetry breaking is more than 10\% and even reach 31\% at the amplitude level in the FAT approach \cite{Zhou:2015jba} compared to the conventional $SU(3)_F$ symmetrical topological diagrammatic approach \cite{Chiang:2007bd}.
Besides, the $\chi^2$ fit in the conventional topological diagrammatic approach
has to be performed for each category of decays $\overline B\to DP$, $\overline B\to D^{*}P$ and $\overline B\to DV$ decays due to the large difference between pseudoscalar and vector mesons,  resulting in three sets of parameters.
With so many parameters, they lost the predictive power of the branching fractions.
Even though with much more parameters than in the FAT approach, its $\chi^2$ per degree of freedom ($d.o.f.$) is larger
($\chi^{2} /d.o.f.=1.4$ in FAT \cite{Zhou:2015jba}).

Compared with the QCD-inspired methods \cite{Beneke:1999br, Bauer:2001cu, Li:2008ts, Zou:2009zza}, the amplitudes of color-suppressed $C$ diagrams are relatively large in the FAT approach where the non-factorizable contribution are dominant, as well as in the topological approach \cite{Chiang:2007bd}.
The hierarchies of topological amplitudes obtained in FAT are \cite{Zhou:2015jba}
\begin{eqnarray}\label{TCE}
|T_c^{DP}|:|C_c^{DP}|:|E_c^{DP}| &\sim & 1:0.45:0.1\, , \nonumber \\
|T_c^{D^*P}| :|C_c^{D^*P}| :|E_c^{D^*P}|&\sim & 1:0.36:0.1\, ,\nonumber\\
|T_c^{DV}|:|C_c^{DV}|:|E_c^{DV}|&\sim &1:0.31:0.1\, ,
\end{eqnarray}
which differ from the relation $|T_c^{DP}|\gg |C_c^{DP}| \sim |E_c^{DP}|$ calculated in the PQCD approach \cite{Li:2008ts}.
The relatively larger $C$ diagrams have significant impacts on the processes without $T$ diagrams. For example, the topological amplitudes of $\overline B^{0}\to D^{0}\rho^{0}$ and $D^{0}\omega$ decays are $(E-C)/\sqrt2$ and $(E+C)/\sqrt2$, respectively. The branching fraction of the $D^{0}\rho^{0}$ mode is predicted to be almost one half of that of the $D^{0}\omega$ mode in the PQCD approach \cite{Li:2008ts}, since $C$ and $E$ diagrams contribute destructively for the former mode but constructively for the latter one, which does not agree with  the experiment.
In the FAT approach, this issue can be explained since both channels
are dominated by the $C$ diagram.
It is easy to see from Eq.(\ref{TCE}) that there is non-negligible
difference for the $C$ contributions between different category of decays
$B \to DP$, $B \to D^*P$ and $B \to D V$.

The relatively large  amplitudes of color-suppressed $C$ diagrams in the FAT approach
can also used to explain the isospin asymmetry.
The $\overline B\to D \pi$ system can   be decomposed in terms of two isospin amplitudes, $A_{1/2}$ and $A_{3/2}$, which correspond to the transition into $D\pi$ final states with isospin $I=1/2$ and $I=3/2$, respectively. The ratio
\begin{equation}\label{iso-rat}
\frac{A_{1/2}}{\sqrt{2} A_{3/2}}=1+{\cal O}(\Lambda_{\rm QCD}/m_b)\, ,
\end{equation}
is a measure of the departure from the heavy-quark limit~\cite{Beneke:2000wa}.
The isospin amplitudes can be expressed by the topological amplitudes as
\begin{equation}\label{iso-rat-calc}
\frac{A_{1/2}}{\sqrt{2}A_{3/2}}=1-\frac{3}{2}\left(\frac{C-E}{T+C} \right)\, ,
\end{equation}
and the numerical result in FAT \cite{Zhou:2015jba} is
\begin{eqnarray}
\left|\frac{A_{1/2}}{\sqrt{2}A_{3/2}}\right|_{D\pi}= 0.65 \pm 0.03,
\end{eqnarray}
together with relative strong phase between the $I=3/2$ and $I=1/2$ amplitudes,
 \begin{eqnarray}
\cos \delta= 0.90\pm 0.04 .
\end{eqnarray}
Comparing with Eq.(\ref{iso-rat}), we observe that the isospin-amplitude ratio shows significant deviation from the heavy-quark limit. Because the  contribution from annihilations has been neglected,  we can trace this feature back to the large color-suppressed $C$ topologies.
\subsection{Two-body charmless $B$ meson decays $B \to PP, PV$}
\begin{figure}[thp!]
 \begin{center}
 \includegraphics[scale=0.6]{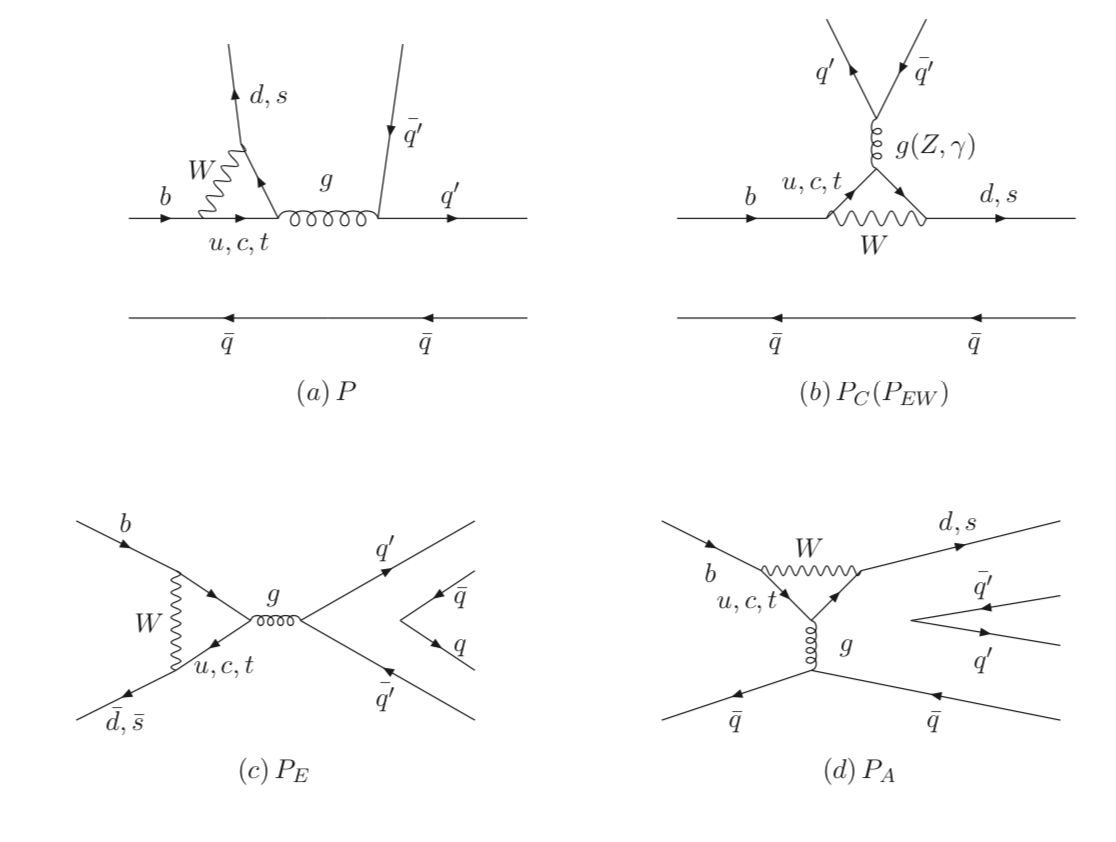}
 \caption{Topological penguin diagrams contributing to
  $B\to PP $ and $B\to PV$ decays:
 (a) the color-favored QCD-penguin diagram, $P$;
 (b) the flavor-singlet QCD-penguin diagram, $P_C$
       and EW-penguin diagram $P_{EW}$;
 (c) the exchange type QCD-penguin diagram, $P_E$ and
 (d) the QCD-penguin annihilation diagram, $P_A$. The diagrams are taken from~\cite{Zhou:2015jba}.}
 \label{Penguin}
 \end{center}
 \end{figure}

Different from two-body charmed $B$ meson decays with only tree diagrams, charmless
$B$ meson decays also receive contributions from penguin diagrams enhanced by CKM matrix elements.
Specifically, corresponding to 4-types of tree diagrams, four types of one loop penguin diagrams grouped into QCD penguin and electro-weak penguin (EW penguin) topologies should be added.
Ignoring some small diagrams, only five penguin diagrams are kept: color-favored QCD penguin emission diagram $P$, color-suppressed QCD penguin emission diagram $P_C$, $W$-annihilation penguin diagram $P_A$, the $W$ penguin exchange diagram $P_E$ and
electro-weak penguin emission diagram $P_{EW}$, as shown in Fig.~\ref{Penguin}.

Similar to the case of the charmed $B$ decays, the non-perturbative contributions of tree diagrams are parameterized as universal magnitudes ($\chi$) and phases ($\phi$).
However, we can not fit three types of charmless $B$ decay amplitude categories $B \to PP$, $B \to PV$ ($P$ emission) and $B \to VP$ ($V$ emission)
together.
For emission diagrams $C$, the case of the pseudo-scalar emission
is different from case of the vector emission. We hence parameterize the $C$ diagram magnitude and associated phase as $\chi^{C}\mathrm{e}^{i\phi^{C}}$ in $B\to PP$, $VP$ decays and
$\chi^{C^{\prime}}\mathrm{e}^{i\phi^{C^{\prime}}}$ in $B\to PV$, respectively,
to distinguish cases in which the emitted meson is pseudo-scalar or vector.
In terms of QCD penguin diagram amplitude, we consider all contributions from every topological diagram in Fig.~\ref{Penguin}, where topology $P$ contributes most.
The leading contribution from topology $P$ diagram is similar to the color favored tree diagram $T$ and predicted from QCD calculations in all the three types of $B \to PP$, $B\to VP$ and $B\to PV$ categories.
Since the chiral enhancement only contributes to the pseudo-scalar meson
(Goldstone boson) emission diagram, we include it only in $B \to PP$  and $B\to VP$ categories by introducing two parameters $\chi^{P}$, $\phi^{P}$.
In the flavor-singlet QCD penguin diagram $P_C$, we distinguish them as $\chi^{P_C}$,
$\phi^{P_C}$ for $B \to PP$ and $B \to VP$ decays and $\chi^{P_C^{\prime}}$,
$ \phi^{P_C^{\prime}}$ for $B \to PV$ decays, respectively.
The penguin annihilation diagram $P_A$ is not distinguishable in weak interaction from the diagram $P$, we include the non-perturbative contributions of $P_A$ diagram in the parameter $\chi^{P}$, $\phi^{P}$ in the $B \to PP$ and $B\to VP$ modes.
While, for $B\to PV$ decay, we introduce two parameters $\chi^{P_A}$, $\mathrm{e}^{i\phi^{P_A}}$
for penguin annihilation diagram $P_A$.
The contribution from $P_E$ is expected to be smaller than $P_A$ and negligible.
For EW-penguin diagrams, we only keep the largest contribution diagram from
EW-penguin contributions, $P_{EW}$.
Similar to the $T$ diagram, we evaluate $P_{EW}$ without introducing new parameters.

The number of free parameters in FAT approach is significantly reduced from the previous
topological diagram approach \cite{Cheng:2014rfa}.
We fit 14 parameters from 37 experimental measured branching fractions and 11 CP asymmetry parameters of $B \to PP$ and $B \to PV$ decays,
with much smaller $\chi^2$ per degree of freedom ($\chi^2/\text{d.o.f}=1.3$ in FAT approach) than in previous topological diagram approach \cite{Cheng:2014rfa}.
The mapping of these topologies to well-known QCDF amplitudes introduced in
\cite{Beneke:2003zv, Beneke:2001ev}, is shown in Table \ref{FATQCDF}.
\begin{table}
\caption{The amplitudes and strong phases of topological diagrams in the FAT approach \cite{Zhou:2015jba} corresponding to contributions in the QCDF~\cite{Beneke:2001ev}.
  The topology $A$ and $P_E$  are neglected in the FAT approach. The electroweak penguin contributions of
 $\alpha_4^{\mathrm{EW}}$,  $\beta_3^{\mathrm{EW}}$ and $\beta_4^{\mathrm{EW}}$ in the QCDF are also neglected in the FAT approach.
}
\begin{center}
\label{FATQCDF}
\scriptsize\begin{tabular}{|c||c|c|c|c|c|c|c|c|c|}
\hline
Diagram  & T  &  C  & $P_C$  & P(PP)
& $P_{EW}$ &  E  &  A  & $P_A$(PV)  &$P_E$  \\
\hline
FAT   & $a_1$  & $\chi^{C^{(\prime)}}\mathrm{e}^{i\phi^{C^{(\prime)}}}$ & $\chi^{P_C^{(\prime)}}\mathrm{e}^{i\phi^{P_C^{(\prime)}}}$ &$a_{4}(\mu)+ \chi^{P}\mathrm{e}^{i\phi^{P}}r_{\chi}$
 &  $a_9(\mu)$ & $\chi^{E} \mathrm{e}^{i\phi^{E}} $  & $-$  & $-i \chi^{P_A}\mathrm{e}^{i\phi^{P_A}}$ &- \\
 &-&$0.48\mathrm{e}^{-1.58i}$  &$0.048\mathrm{e}^{1.56i}$&$-0.12\mathrm{e}^{-0.24i}$&-0.009&$0.057\mathrm{e}^{2.71i}$&&
 $0.0059\mathrm{e}^{-0.006i}$&\\
 \hline
QCDF & $\alpha_1$ &$\alpha_2$ &$\alpha_3$ & $\alpha_4$ & $\alpha_3^{\mathrm{EW}}$ &$\beta_1$ &$\beta_2$ &$\beta_3$& $\beta_4$\\
&-&$0.22\mathrm{e}^{-0.53i}$  &$0.011\mathrm{e}^{2.23i}$&$-0.089\mathrm{e}^{0.11i}$&$-0.009 \mathrm{e}^{0.04i}$&0.025&-0.011
&-0.008&-0.003\\
 \hline
\end{tabular}
\end{center}
\end{table}
It is apparent that there are huge differences between numbers fitted from experimental data in the FAT and the calculated results in the QCDF \cite{Beneke:2001ev}, especially for the strong phases.
That might be the reason why the current QCDF and SCET calculation
can predict the branching ratios but can not explain the direct CP asymmetries well. Apart form QCD corrections \cite{Huber:2015bva,Bell:2015koa,Huber:2016xod,Huber:2020pqb,Bell:2020qus,Li:2012nk,Cheng:2014fwa,Cheng:2015qra,Cheng:2018khi}, power corrections
in QCD calculations \cite{Beneke:2018wjp,Wang:2017ijn,Wang:2018wfj,Lu:2018cfc,Gao:2019lta,Li:2020rcg,Shen:2018abs,Shen:2019vdc,Shen:2019zvh}
and also QED corrections \cite{Beneke:2020vnb,Beneke:2021jhp} might be crucial to solve such problems.

\begin{table}
\caption{The direct CP asymmetries ($\mathcal{A}$) and
mixing-induced CP asymmetries   ($\mathcal{S}$)
of $\overline B\to PP$ decays predicted by the FAT approach~\cite{Zhou:2015jba} compared to data~\cite{ParticleDataGroup:2020ssz} and the results from
the conventional flavor diagram approach~\cite{Cheng:2014rfa}.}\label{tab:CPBPP}
\begin{center}
\begin{tabular}{cccccccc}
\hline\hline
  & Mode &  ~~~~~$\mathcal{A_{\rm exp}}$~~ & ~~$\mathcal{A}_{\text{FAT}}$~~~~~~&$\mathcal{A}_{\text{Flavor~diagram}}$~~~~&
~~~~~~~~$\mathcal{S_{\rm exp}}$~~&~~~$\mathcal{S}_{\text{FAT}}$~~&~~$\mathcal{S}_{\text{Flavor~diagram}}$~~\\
\hline
&$ \pi^{+} \pi^{-}$     &$0.32\pm0.04$   &$0.31\pm0.04$     &  $0.326\pm0.081$    &$-0.65\pm0.04$&   $-0.60\pm0.03$   &    $-0.717\pm0.061$       \\
&$  \pi^{0}  \pi^{0}$    &$0.33\pm0.22$         &$0.57\pm0.06$        &  $0.611\pm0.113$      &$     $&        $0.58\pm0.06$  &      $0.454\pm0.112$       \\
&$  \pi^{0} \eta$        &$     $               &$-0.16\pm0.16$       &  $0.566\pm0.114$      &$     $&        $-0.98\pm0.04$&       $-0.098\pm0.338$       \\
&$  \pi^{0} \eta^{'}$    &$     $               &$0.39\pm0.14$        &  $0.385\pm0.114$      &$     $&        $-0.90\pm0.07$ &      $0.142\pm0.234$       \\
&$ \eta \eta$            &$     $               &$-0.85\pm0.06$       &   $-0.405\pm0.129$   &$     $&         $0.33\pm0.12$ &       $-0.796\pm0.077$       \\
&$ \eta \eta^{'}$        &$     $               &$-0.97\pm0.04$       &   $-0.394\pm0.117$   &$     $&         $-0.20\pm0.15$ &      $-0.903\pm0.049$       \\
&$ \eta^{'} \eta^{'}$    &$     $               &$-0.87\pm0.07$       &   $-0.122\pm0.136$   &$     $&         $-0.46\pm0.14$ &      $-0.964\pm0.037$       \\
&$  \pi^{0}  K_s$ &$0.00\pm0.13$     &$-0.14\pm0.03$    &  $-0.173\pm0.019$    &$0.58\pm0.17$ &    $0.73\pm0.01$  &      $0.754\pm0.014$         \\
&$ \eta  K_s$     &$     $               &$-0.30\pm0.10$       &  $-0.301\pm0.041$  &$     $         &  $0.68\pm0.04$  &      $0.592\pm0.035$         \\
&$ \eta^{'}  K_s$ &$0.06\pm0.04$     &$0.030\pm0.004$   &  $0.022\pm0.006$   &$ 0.63\pm0.06$   &   $0.69\pm0.00$  &      $0.685\pm0.004$          \\
&$ K^{0} \bar{K^{0}}$    &                      &$-0.057\pm0.002$     & $0.017\pm0.041$     &$0.8\pm0.5$     & $0.099\pm0.002$   &         0      \\
\hline
&$ \pi^{-}\pi^{0}$  &$0.03\pm0.04$    &$-0.026\pm0.003$     &  $0.069\pm0.027$      &$     $&           $ $   &            \\
&$ \pi^{-} \eta$         &$-0.14\pm0.07$                     &$-0.14\pm0.07$       &  $-0.081\pm0.074$     &$     $&           $ $   &           \\
&$ \pi^{-} \eta^{'}$     &$0.06\pm0.16$                      &$0.37\pm0.07$        &  $0.374\pm0.087$      &$     $&           $ $   &           \\
&$ \pi^{-} \bar{K^{0}}$  &$-0.017\pm0.016$                   &$0.0027\pm0.0001$        &   0                   &$     $         &$ $   &           \\
&$  \pi^{0} K^{-}$       &$0.037\pm0.021$                    &$0.065\pm0.024$          &   $0.047\pm0.025$    &$     $         &$ $   &           \\
&$ \eta K^{-}$           &$-0.37\pm0.08$               &$-0.22\pm0.08$     &   $-0.426\pm0.043$   &$     $         &$ $   &           \\
&$ \eta^{'} K^{-}$       &$0.04\pm0.011$                    &$-0.021\pm0.007$         &   $-0.027\pm0.008$   &$     $         &$ $   &           \\
&$ K^{-} K^0$          &$-0.21\pm0.14$                       &$-0.057\pm0.002$        &       0               &$     $         &$$   &           \\
&$ \pi^{+} K^{-}$        &$ -0.083\pm0.004$             &$-0.081\pm0.005$   &   $-0.080\pm0.011$   &$     $         &$$   &           \\
\hline\hline
\end{tabular}
\end{center}
\end{table}

We list the direct CP asymmetries ($\mathcal{A}$) and mixing-induced CP asymmetries
($\mathcal{S}$) of $\overline B\to PP$ decays in Table \ref{tab:CPBPP}, with the addition of the results from conventional flavor diagram approach \cite{Cheng:2014rfa} for comparison.
The $B^{-} \to \pi^{0} {K^{-}}$ decay is naively expected to have the same dominant decay amplitude $T$ and $P$ as $\overline B^{0}\to \pi^{+} {K^{-}}$ decay, and thus one expects similar direct CP asymmetries \cite{Cheng:2009cn}.
However, experimentally these two direct CP asymmetries are quite different, even with an opposite sign. That is the so-called $\pi K$ CP-puzzle.
From the Table \ref{FATQCDF}, we know
the large $C$ contribution with large strong phase can resolve the so called $\pi K$ puzzle.
However, only a large $C$ magnitude can not explain another puzzle, the $\pi \pi$ puzzle:
theoretically $\mathcal{B}(B^0\to \pi^0 \pi^0) < \mathcal{B}(B^0\to \pi^0 \rho^0) < \mathcal{B}(B^0 \to \rho^0 \rho^0) $,
but experimentally they are in the inverse order.
In the FAT approach, the $\pi \pi$ puzzle can be resolved by different strong phase $\phi^{C}$ and $\phi^{C^{\prime}}$ representing the pseudo-scalar and vector meson emission, respectively, even though $\chi^C$ and $\chi^{C^{\prime}}$ are in similar size \cite{Zhou:2015jba}:
\begin{align}\label{parameter}
\chi^{C}=0.48 \pm 0.06,~~~&\phi^{C}=-1.58 \pm 0.08,\nonumber \\
\chi^{C^{\prime}}=0.42 \pm 0.16,~~~&\phi^{C^{\prime}}=1.59\pm 0.17\, .
\end{align}
The difference in strong phase is also agreement with the Glauber phase factor \cite{Li:2014haa}, associated with the Goldstone boson $\pi$, to resolve
the $B\to\pi\pi$, $B\to \pi\rho$ and $B \to \rho\rho$ puzzles consistently.

\subsection{$B$ decays with two vector meson $B\to VV$}
The vector mesons can be produced in three polarization states, corresponding to the longitudinal $L$ and two helicity $\pm 1$ states. So the decay amplitudes can be described with definite final state helicity $\mathcal{A}_{L,\pm}$.
These decays have rich polarization observables, apart from decay widths and CP asymmetries in contrast to the $PP(PV)$ final states. Theoretically, the longitudinal polarisation amplitude ${\mathcal A}_L$  is similar to the decay in $PP,PV$ final states. Then we have $8$ unknown real parameters to be fitted for ${\mathcal A}_L$.
For the transverse amplitudes, it is convenient to use the linear polarised form ${\mathcal A}_{\parallel}=({\mathcal A}^++{\mathcal A}^-)/\sqrt 2$, ${\mathcal A}_{\perp}=({\mathcal A}^+-{\mathcal A}^-)/\sqrt 2$. The transverse amplitudes are power suppressed relative to longitudinal amplitude.
So the endpoint divergences of transverse amplitudes emerge even at leading power in QCDF \cite{Beneke:2006hg}, which decreases the predictive power.
In the FAT approach, besides the color-favored tree diagram $T$ and QCD-penguin diagram $P$ which do not introduce any parameter, we only consider the panguin annihilation diagram $P_A$.
Thus, there are only 10 universal parameters totally, which will be fitted by experimental data. The best-fitted values of the parameters are given in \cite{Wang:2017hxe}.

Because of the $V-A$ coupling of weak interaction, a specific pattern of the three helicity amplitudes is naively expected~\cite{Koerner1979}:
\begin{eqnarray}
	{\bar{\mathcal A}}^0:\bar{\mathcal A}^-:\bar{\mathcal A}^+=1:\frac{\Lambda_{QCD}}{m_b}:\left(\frac{\Lambda_{QCD}}{m_b}\right)^2\,,
	\label{herachy}
\end{eqnarray}
for $\overline{B}$ meson.
For $B$ meson, we exchange the superscript $-$ and $+$. The amplitudes of tree dominated decays respect the hierarchy (\ref{herachy}), and $f_L$ are closed to $1$.
Taking the three $B\to\rho\rho$ decay modes in Table \ref{tree} as an example,
we list the numerical  results of longitudinal polarization for each topological diagram of these decays,
\begin{eqnarray}
	|T^{B\to\rho_L\rho_L}|:|C^{B\to\rho_L\rho_L}|:|E^{B\to\rho_L\rho_L}|:|P^{B\to\rho_L\rho_L}|:|P_A^{B\to\rho_L\rho_L}|=1:0.22:0.21:0.14:0.08.
\end{eqnarray}
For the decay  $B^-\to\rho^- \rho^0$, although the absolute value of the $C$ diagram is suppressed,  it can enhance the  magnitude of the decay amplitude by $20\%$.
With the larger contribution from $C$ diagram, the large branching fraction of $\overline B^0\to\rho^0 \rho^0$ is also explained.
\begin{table}[h]\centering
	\caption{Comparison of the FAT approach \cite{Wang:2017hxe} and the experimental results \cite{Zyla:2020zbs} for observables of $B^-\to \rho^- \rho^0$, $\overline{B}^0\to \rho^+ \rho^-$ and $\overline{B}^0\to \rho^0\rho^0$.}	\label{tree}
	\begin{tabular}{cccccc}
		\hline
		\hline
		Mode & $\mathcal{B}(10^{-6})$ & $f_L(\%)$ & $f_\perp(\%)$ & $\phi_\parallel$ (rad) & $\phi_\perp$ (rad)\\
		\hline
		$B^-\to \rho^- \rho^0$  & $21.7\pm5.1$ & $95.5\pm1.5$ & $2.22\pm0.64$ & $-0.09\pm0.05$ & $-0.09\pm0.05$\\
		Expt. &  $24.0\pm1.9$ & $95\pm1.6$ &  &  & \\
		\hline
		$\overline{B}_0\to \rho^+ \rho^-$ &  $29.5\pm6.5$ & $92.6\pm1.6$ & $3.65\pm0.91$ & $-0.27\pm0.08$ & $-0.27\pm0.08$\\
		Expt. &  $27.7\pm1.9$ & $99.0_{-1.9}^{+2.1}$ &  &  & \\
		\hline
		$\overline{B}^0\to \rho^0 \rho^0$ &  $0.94\pm0.49$ & $81.7\pm10.8$ & $9.21\pm5.50$ & $-0.04\pm0.44$ & $-0.03\pm0.44$\\
		Expt. &  $0.96\pm0.15$ & $71_{-9}^{+8}$&  &  & \\
		\hline
		\hline
		& $A_{CP}(\%)$ & $A_{CP}^0(\%)$ & $A_{CP}^\perp(\%)$ & $\Delta \phi_\parallel$ (rad) & $\Delta \phi_\perp$ (rad)\\
		\hline
		$B^-\to \phi K^{*-}$  & $0$ & $0$ & $0$ &  $0$ &  $0$\\
		Expt. &  $-5\pm5$ &  &  &  & \\
		\hline
		$\overline{B}_0\to \phi \overline{K}^{*0}$  & $-8.10\pm2.94$ & $1.30\pm0.54$ & $-16.3\pm8.2$ & $-0.41\pm0.05$ & $-0.41\pm0.05$\\
		\hline
		$\overline{B}_s\to \phi \phi$  & $49.7\pm13.4$ & $10.5\pm9.6$ & $-46.9\pm13.9$ & $1.89\pm0.19$ & $1.89\pm0.19$\\
		\hline
		\hline
	\end{tabular}
\end{table}

However, large transverse polarization fractions ($\sim 50\%$) of  $B\to \phi K^*$  have been measured by Babar \cite{BaBar:2003spf} and Belle \cite{Belle:2003ike} in 2003.
Later on, some other penguin-dominated decays, such as $B\to \rho K^*$ and $B_s\to \phi \phi$, have also been found with large transverse polarization fractions. As mentioned before, we use the $P_A$ diagram to explain these large transverse polarization fraction. In Table \ref{penguion}, we list the results of three best measured channels  $B^-\to \phi K^{*-}$, $\overline{B}^0\to \phi \overline{K}^{*0}$ and $\overline{B}_s\to \phi\phi$ as an illustration.
In these decays, the magnitudes of QCD penguin diagram $P,P_c$,  and  $P_A$ are at the same order.
For $B_s \to \phi\phi$ decay, our center value is a bit larger than the experimental data, and the predicted polarization fractions are in agreement with the data.
The acceptable divergency is in fact constrained by the measured branching ratio of $B_s \to \phi K^*$ decay.
\begin{table}[h]\centering
	\caption{Comparison of the FAT approach \cite{Wang:2017hxe} and the experimental results \cite{Zyla:2020zbs} for observables of $B^-\to \phi K^{*-}$, $\overline{B}^0\to \phi \overline{K}^{*0}$ and $\overline{B}_s\to \phi\phi$.}\label{penguion}
	\begin{tabular}{cccccc}
		\hline
		\hline
		Mode & $\mathcal{B}(10^{-6})$ & $f_L(\%)$ & $f_\perp(\%)$ & $\phi_\parallel$ (rad) & $\phi_\perp$ (rad)\\
		\hline
		$B^-\to \phi K^{*-}$  & $9.31\pm2.81$ & $48.0\pm16.0$ & $25.9\pm8.6$ & $2.47\pm0.27$ & $2.47\pm0.27$\\
		Expt. &  $10\pm2$ & $50\pm5$ & $20\pm5$ & $2.34\pm0.18$ & $2.58\pm0.17$\\
		\hline
		$\overline{B}_0\to \phi \overline{K}^{*0}$ &  $8.64\pm2.61$ & $48.0\pm16.0$ & $26.0\pm8.6$ & $2.47\pm0.27$ & $2.47\pm0.27$\\
		Expt. &  $10.0\pm0.5$ & $49.7\pm1.7$ & $22.4\pm1.5$ & $2.43\pm0.11$ & $2.53\pm0.09$\\
		\hline
		$\overline{B}_s\to \phi \phi$ &  $26.4\pm7.6$ & $39.7\pm16.0$ & $31.2\pm8.9$ & $2.53\pm0.28$ & $2.56\pm0.27$\\
		Expt. &  $18.7\pm1.5$ & $37.8\pm1.3$ & $29.2\pm0.9$ & $2.56\pm0.06$ & $2.818\pm0.192$\\
		\hline
		\hline
		& $A_{CP}(\%)$ & $A_{CP}^0(\%)$ & $A_{CP}^\perp(\%)$ & $\Delta \phi_\parallel$ (rad) & $\Delta \phi_\perp$ (rad)\\
		\hline
		$B^-\to \phi K^{*-}$  & $1.00\pm0.27$ & $1.26\pm0.71$ & $-1.16\pm0.30$ & $-0.02\pm0.00$ & $-0.02\pm0.00$\\
		Expt. &  $-1\pm8$ & $17\pm11$ & $22\pm25$ & $0.07\pm0.21$ & $0.19\pm0.21$\\
		\hline
		$\overline{B}_0\to \phi \overline{K}^{*0}$  & $1.00\pm0.27$ & $1.26\pm0.71$ & $-1.16\pm0.30$ & $-0.02\pm0.00$ & $-0.02\pm0.00$\\
		Expt. &  $0\pm4$ & $-0.7\pm3.0$ & $-2\pm6$ & $0.05\pm0.05$ & $0.08\pm0.05$\\
		\hline
		$\overline{B}_s\to \phi \phi$  & $0.83\pm0.28$ & $1.55\pm0.85$ & $-1.02\pm0.29$ & $-0.01\pm0.00$ & $-0.01\pm0.00$\\
		\hline
		\hline
	\end{tabular}
\end{table}

\section{Summary and outlook}\label{sum}

The FAT approach is a data driven framework to study heavy meson decays.
Compared to QCD calculations, it gets a better chance to be consistent with data, if all crucial dynamics are figured out and parameterized in a proper way.
It is useful in two aspects, one experimentally and the other one theoretically.
From the experimental point of view, it can make trustworthy predictions for unmeasured observables, {\it e.g.}, $D/B$ meson decay branching ratios and also CP asymmetries, which will help the discovery of new decay channels and new CP violation effects in experiments.
One highlight is the prediction for $\Delta A_{CP}$ confirmed by later LHCb measurement,
which was also the first direct CP violation discovery in the charm sector.
From the theoretical point of view, the amplitude structures of the decay processes can be decoded in the FAT approach, which will provide hints for the directions of future theoretical studies.
For example, the understanding of the large magnitude and large strong phase in the $C$ amplitude in $B$ decays require more theoretical efforts.

In the end, it is important to clarify that the FAT approach should not refer to any of the parametrization schemes that has appeared in any of the existing papers \cite{Li:2012cfa,Qin:2013tje,Zhou:2015jba,Zhou:2016jkv}.
Apparently, the parametrization schemes for $D$ and $B$ decays are quite different from each other.
Instead, it should be regarded as a more general framework, in which the short distance dynamics and the long distance dynamics of different topological amplitudes are factorized with the former formulated by corresponding Wilson coefficients and the latter as hadronic matrix elements either calculable or parameterized.
In the parametrization, important SU(3) breaking effects are expected to be figured out, which leads to some arbitrariness and also leaves space to improve.
Until now, although the FAT approach have made many remarkable achievements, is still await more efforts in the future to improve the parametrization schemes for both the $D$ and $B$ decays, especially in such a stage with more and more precise data.

\begin{acknowledgements}

The authors would like to thank Professors Xin Liu, Zhen-Jun Xiao and Ruilin Zhu for the invitation to write a review article on the factorization-assisted topological-amplitude approach
in the special issue “Heavy Flavor Physics and CP Violation” of Advances in High Energy Physics.
The authors are grateful to Hsiang-nan Li, Cai-Dian L\"u and Fu-Sheng Yu for original works in innovating the FAT approach.
This work is supported by Natural Science Foundation of China under grant Nos. 12005068, 12105148, 12105112, 12105099
and the Natural Science Foundation of Jiangsu Education Committee with Grant No. 21KJB140027.

\end{acknowledgements}

\end{document}